\documentclass{article}
\usepackage{spconf,amsmath,graphicx}
\usepackage{enumitem}
\setlist{nosep, leftmargin=14pt}
\usepackage{mwe} % to get dummy images
\usepackage{amsmath}
\usepackage{amsfonts}
\RequirePackage{bibspacing}

\title{A Novel Deep Learning Tractography Fiber Clustering Framework\\ for Functionally Consistent White Matter Parcellation\\ Using Multimodal Diffusion MRI and Functional MRI}
\name{\parbox{\linewidth}{\begin{center}Jin Wang$^{1,*}$, Bocheng Guo$^{1,*}$, Yijie Li$^{1}$, Junyi Wang$^{1}$  Yuqian Chen$^{2}$, Jarrett Rushmore$^{2,3}$,\\\textit{Nikos Makris}$^{2,3}$, \textit{Yogesh Rathi}$^{2}$, \textit{Lauren J O’Donnell}$^{2,\#}$, \textit{Fan Zhang}$^{1,\#}$ \end{center}}
\thanks {*Jin Wang and Bocheng Guo are co-first-authors; \#Corresponding to co-authors Fan Zhang (fan.zhang@uestc.edu.cn) and Lauren J .O’Donnell(odonnell@bwh.harvard.edu).}}
\address{$^{1}$University of Electronic Science and Technology of China, Chengdu, China\\$^{2}$ Harvard Medical School, Boston, USA\\$^{3}$ Boston University, Boston, USA}
\begin{document}

\maketitle
\begin{abstract}
Tractography fiber clustering using diffusion MRI (dMRI) is a crucial strategy for white matter (WM) parcellation. Current methods primarily use the geometric information of fibers (i.e., the spatial trajectories) to group similar fibers into clusters,  overlooking the important functional signals present along the fiber tracts. There is increasing evidence that neural activity in the WM can be measured using functional MRI (fMRI), offering potentially valuable multimodal information for fiber clustering. In this paper, we develop a novel deep learning fiber clustering framework, namely \emph{Deep Multi-view Fiber Clustering (DMVFC)}, that uses joint dMRI and fMRI data to enable functionally consistent WM parcellation. DMVFC can effectively integrate the geometric characteristics of the WM fibers with the fMRI BOLD signals along the fiber tracts. It includes two major components: 1) a multi-view pretraining module to compute embedding features from fiber geometric information and functional signals separately, and 2) a collaborative fine-tuning module to simultaneously refine the two kinds of embeddings. In the experiments, we compare DMVFC with two state-of-the-art fiber clustering methods and demonstrate superior performance in achieving functionally meaningful and consistent WM parcellation results.
\end{abstract}
\section{Introduction}
\label{sec:intro}

Diffusion magnetic resonance imaging (dMRI) tractography \cite{basser2000vivo} is a well-established neuroimaging technique that uniquely allows \textit{in vivo} mapping of the brain's white matter (WM) connections at the macroscopic level. This technique has been widely used in the quantitative analysis of the brain's structural connectivity \cite{Zhang2022-kb}. Fiber clustering is a crucial strategy for WM parcellation to subdivide whole-brain tractography into geometrically similar and physiologically meaningful bundles \cite{Garyfallidis2012-io,Zhang2018-vq,O-Donnell2007-qh}. However, almost all existing clustering methods suffer from a common limitation that they do not explicitly capture the functional implication of fiber clusters.

Functional magnetic resonance imaging (fMRI) based on blood oxygen level-dependent (BOLD) contrast \cite{ding2018detection} is a mature technology for measuring the functional activities of the brain. Traditionally, fMRI signals have been thought to originate from postsynaptic potentials in the gray matter (GM), which are essentially absent in the WM. Recent studies, however, have demonstrated that fMRI signals in WM reliably respond to stimuli in patterns similar to those observed in GM \cite{courtemanche2018detecting,fraser2012white}. Notably, the resting-state frequency spectrum of WM fMRI signals closely parallels that of GM. Additionally, stable long-range functional networks can be identified in WM at the voxel level \cite{peer2017evidence}, and these networks are well correlated with those in GM, which indicates functional homogeneities of WM fMRI signals within fiber tracts \cite{ding2018detection}. Therefore, in addition to the spatial location of WM fibers, a promising solution to enable functionally consistent WM parcellation is to include fMRI signals for joint multimodal dMRI and fMRI processing.

Deep neural networks have demonstrated superior performance in various computational neuroimaging tasks such as registration \cite{DDMReg} and segmentation \cite{Wang2023-xh}. In particular, deep-learning-based clustering is a popular unsupervised learning tool \cite{Ren2024-ca}, which can be used for fiber clustering. One straightforward approach to unsupervised deep clustering is to extract feature embeddings with neural networks and then perform clustering on these embeddings to form clusters \cite{chen2023deep}. The learned embeddings are high-level representations of input data and have been demonstrated to be informative for downstream tasks. For fiber clustering, one promising approach for learning feature embeddings is self-supervised learning, which has been proven to be an efficient method and shows advanced performance in many applications \cite{Xie2023-pn}. Furthermore, multi-view clustering has recently been the focus of attention through the use of information from multiple views \cite{Yang2018-kw}. The key to multi-view clustering is effectively mining information from multiple views for better clustering performance, which can help us combine the multimodal information of fMRI with the fiber geometric information.

In this paper, we develop a novel deep learning fiber clustering framework, namely \emph{Deep Multi-view Fiber Clustering (DMVFC)}, that uses joint dMRI and fMRI data to enable functionally consistent WM parcellation. DMVFC can effectively integrate the geometric characteristics of the WM fibers with the fMRI BOLD signals along the fiber tracts. It includes two major components: 1) a multi-view pretraining module to compute embedding features from fiber geometric information and functional signals separately, and 2) a collaborative fine-tuning module to simultaneously refine the two kinds of embeddings. There are three main contributions:
\vspace{0.5em}
\begin{enumerate}
  \item To our knowledge, this is the first work that leverages deep multi-view clustering for fiber clustering. 
  \item DMVFC uniquely incorporates fMRI data for fiber clustering to enable functionally consistent WM parcellation. 
  \item By integrating different data modalities, our work provides more insights and establishes a new framework for fiber clustering techniques.
\end{enumerate}

\section{Methods}
\label{sec:method}

The overall workflow of DMVFC is shown in Figure 1. There are two main training stages in DMVFC: (1) multi-view pretraining and (2) collaborative fine-tuning. In the first stage, the multi-view pretraining module computes two kinds of embeddings from fiber geometric information and brain functional signals (Section 2.1). In the second stage, the collaborative fine-tuning module refines the pretrained embeddings to ensure that the clustering outcomes integrate both geometric and functional information simultaneously (Section 2.2). 

\begin{figure}

\begin{minipage}[b]{1.0\linewidth}
  \centering
  \centerline{\includegraphics[width=8.5cm]{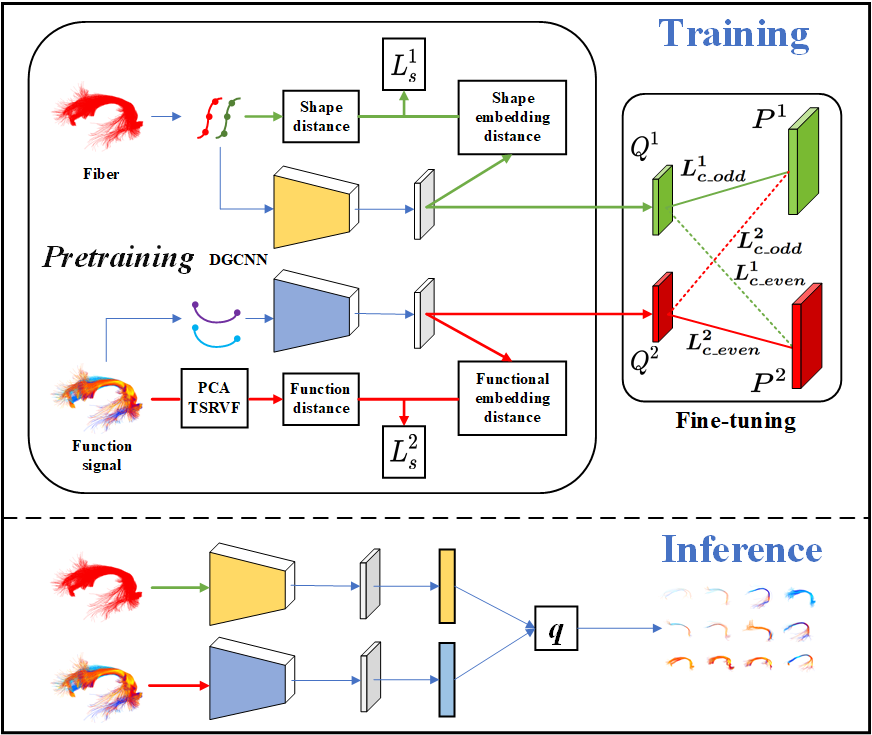}}
%  \vspace{2.0cm}
\end{minipage}
\caption{Overview of DMVFC.}
\label{fig:res}
\end{figure}

\subsection{Multi-view pre-training for embedding computation}
\label{ssec:mvpretraining fot embedding}

In multi-view clustering analysis, the views refer to different representations or perspectives of the same data, allowing the clustering model to leverage complementary information from each view to improve clustering accuracy and robustness \cite{chen2020multi}. In our study, the multi-view pre-training module includes two parallel embedding extraction models to obtain feature embeddings from two views, i.e., the fiber geometry information and brain functional signals, respectively. The two models are based on the popular dynamic graph convolutional neural network (DGCNN) \cite{Wang2019-xc}, which is specifically designed for processing graph-structured data such as point clouds. DGCNN has been shown successful for fiber clustering in the Deep Fiber Clustering (DFC) framework \cite{chen2023deep} that relies on geometry similarity between fibers to compute embedding features and subsequently fine-tunes the embedding space with clustering results. In the proposed DMVFC, we extend this approach to a multi-view model that incorporates both fiber geometry and functional signals.

\subsubsection{Network inputs}
\label{sssec}

To capture geometric information, fibers are represented as point clouds that are defined by the spatial coordinates of fiber points along the trajectories. In this case, for a fiber i, its geometric input is formed as $\mathbf{x}_{i}^{1} \in \mathbb{R} ^{n_p\times 3}$, which represents $n_{p}$ evenly sampled points along the fiber in spatial Right-Anterior-Superior (RAS) coordinates. To capture functional signals, fibers are represented as point clouds defined by the fMRI signals at the endpoints. The fMRI signals at the endpoints of fiber contain 1200 time points per endpoint (see Section 3.1), which were randomly downsampled to 600 for efficient CPU computation. As a result, for a fiber i, its functional signal input is formed as $\mathbf{x}_{i}^2 \in \mathbb{R}^{2\times 600}$.

\subsubsection{Network training}
\label{sssec:mvpretraining fot embedding}

To extract geometric and functional signal embeddings that make each fiber embedding distinguishable and enhance clustering performance, we follow the design principles of the DFC method \cite{chen2023deep}. DFC is a self-supervised learning-based method that takes a pair of fibers as inputs and utilizes the fiber distances as pseudo-labels to train a latent embedding feature for each fiber using DGCNN. In our method, in contrast to the single model used in DFC, we construct two parallel DGCNN models for the aforementioned geometric and functional inputs.

Specifically, $\left\{ (\mathbf{x}_i^v, \mathbf{y}_i^v) \mid \mathbf{x}_i^v, \mathbf{y}_i^v \in \mathbb{R}^{n_{p}\times C} \right\}_{i=1}^N$ denote input pairs fed into the corresponding DGCNN network, $N$ represents the number of fibers, $n_{p}$ indicates the number of points on each data. $C$ is the number of channels for each data type. For geometric input ($v$=1), pseudo-label $S_{i}^{1}$ is calculated as the minimum average direct-flip distance, which is widely applied in WM fiber clustering \cite{Garyfallidis2012-io,Zhang2018-vq}. For functional input ($v$=2), pseudo-label $S_{i}^{2}$ is calculated using the Mean Square Error (MSE) of fMRI signals. In brief, to reduce the computational complexity, the original fMRI signals at each fiber location contain 1200 time points, which are processed by a principal component analysis (PCA) to keep the top 30 principal components. To quantify differences between fMRI signals along fibers, we employ the transport square-root velocity function (TSRVF), transforming the raw data into Riemannian space, as applied in \cite{zhao2023riemannian}. This method is highly discriminative among different fiber clusters, as it reduces within-class variability through alignment and enables low-dimensional encoding of high-dimensional data.

Finally, to ensure that the distance between embeddings aligns with the similarity of data pairs, the loss function used in our study is:
\begin{equation}
L_{s} = {\textstyle \sum_{i=1}^{N}} \left \| d(\mathbf{x}_{i}^{v}, \mathbf{y}_{i}^{v})-S_{i}^{v}  \right \| _{2}^{2} 
\label{equal1}
\end{equation}
Where $S_{i}^{v}$ is the label of input pairs, $d(\cdot)$ is the Euclidean distance between the learned deep embeddings $f_{\theta }^{v} \left ( \mathbf{x}_{i}^{v} \right )$.

\subsection{Collaborative fine-tuning}
\label{ssec:mvpretraining}

As geometric and functional embeddings of fibers are computed individually, it is essential to integrate the complementary information to ensure that clustering predictions are consistent across both fiber geometry and functional signals. To achieve this, we draw inspiration from the deep embedded multi-view clustering with collaborative training (DEMVC) method \cite{xu2021deep}, which uses embeddings as different views to guide the fine-tuning of each other. In this case, this collaborative approach allows each embedding to incorporate complementary information from other views during the fine-tuning process.

In our collaborative fine-tuning stage, we use the loss represented as the weighted sum of $L_{s}$ (see Eq. \ref{equal1}) to ensure that the embedding maintains the similarity between different fibers, and $L_{c}$ with two views in odd and even epochs are defined as follows: 
\begin{equation}
  L_{c\underline{~}odd}^{v}=KL\left ( P^{1}|| Q^{v} \right ) = {\textstyle \sum_{i=1}^{N}} {\textstyle \sum_{j=1}^{K}}p_{ij}^{1}\log_{}{\frac{p_{ij}^{1} }{q_{ij}^{v} } }  
\end{equation}
\begin{equation}
L_{c\underline{~}even}^{v}=KL\left ( P^{2}|| Q^{v} \right ) = {\textstyle \sum_{i=1}^{N}} {\textstyle \sum_{j=1}^{K}}p_{ij}^{2}\log_{}{\frac{p_{ij}^{2} }{q_{ij}^{v} } } 
\end{equation}

\noindent where $q_{ij}^{v}$ represents the probability that one input belongs to a given cluster $j$, which is defined by Student's t-distribution:
\begin{equation}
q_{ij}=\frac{(1+\left \| z_{i}-\mu _{j}  \right \|^{2})^{-1} }{\sum_{j'}(1+\left \| z_{i}-\mu _{j'} \right \| ^{2})^{-1} }
\end{equation}

\noindent and $p_{ij}^v$ is a distribution defined as:
\begin{equation}
p_{ij}^v = \frac{(q_{ij}^v)^2 / \sum_i q_{ij}^v}{\sum_j \left( (q_{ij}^v)^2 / \sum_i q_{ij}^v \right)}
\end{equation}

During the training process, the clustering losses are applied alternately in odd and even epochs. This alternating strategy helps the model learn from different perspectives and prevents it from overly relying on specific patterns. The fine-tuning Loss is defined as :
\begin{equation}
L_{f}= L_{s}+\gamma L_{c}
\end{equation}
Where $\gamma$ is empirically set to be  0.1.

\subsection{Inference stage}
\label{ssec:Instage}
In the inference stage, given a new subject, the geometric and functional inputs are formed (as introduced in Section 2.1.1) and inputted to the network to compute the probabilities of the two inputs belonging to a given cluster. The final prediction is determined by averaging the probabilities across both views.

%In inference stage, we put fMRI fibers and dMRI fibers both into the network. And obtain $q_{ij}^{v}$ , epresenting  the possibility that every fiber belongs to a given cluster. We get the final prediction by the average possibility in both fMRI and dMRI fiber as defined in \cite{xu2021deep}: $$s_{i}=\arg\max~_{j} ~(\frac{1}{V} {\textstyle \sum_{v=1}^{V}}q_{ij}^{v})$$ Where $i$ represents the index of fiber, $j$ represents a given cluster.

\section{EXPERIMENTS AND RESULTS}
\label{sec:pagestyle}

\subsection{Data Acquisition and Preprocessing}
\label{ssec:dataset}

We utilize the well-preprocessed dMRI and resting state fMRI (rsfMRI) data from the unrelated 100 subjects in the Human Connectome Project Young Adult (HCP-YA) \cite{Van-Essen2013-gr}. For each subject, one hour rsfMRI data was acquired on a 3T Tesla scanner, with a spatial resolution of 2×2×2mm$^3$, TR=720ms, TE=33.1ms, and dMRI data was obtained with a spatial resolution of 1.25×1.25×1.25mm$^3$, TR=5520 ms, and TE=89.5ms. 

The overall flow of the data processing steps is shown in Figure 2. The dMRI data is processed through a popular tract segmentation algorithm TractSeg \cite{wasserthal2018tractseg} to identify anatomical fiber bundles, which will be further clustered into smaller clusters. We randomly select 7 fiber bundles for demonstration, including: left and right SLF-I, left and right SLF-II, CC-2, CC-3, and CC-4. The rfMRI data is preprocessed through HCP fMRI minimal pipeline \cite{glasser2013minimal}. Then they were cleaned of spatially specific structured noise (ICA-FIX) \cite{griffanti2014ica} and precisely aligned across subjects by multimodal cortical surface registration (MSMAll). To avoid mixing signals and improve the signal-to-noise ratio and fMRI sensitivity in white matter, we carry out spatial smoothing (4 mm full-width half-maximum, isotropic) for GM and WM separately. These preprocessing steps are done using the SPM12 toolbox. 

\begin{figure}

\begin{minipage}[b]{1.0\linewidth}
  \centering
  \centerline{\includegraphics[width=8.5cm]{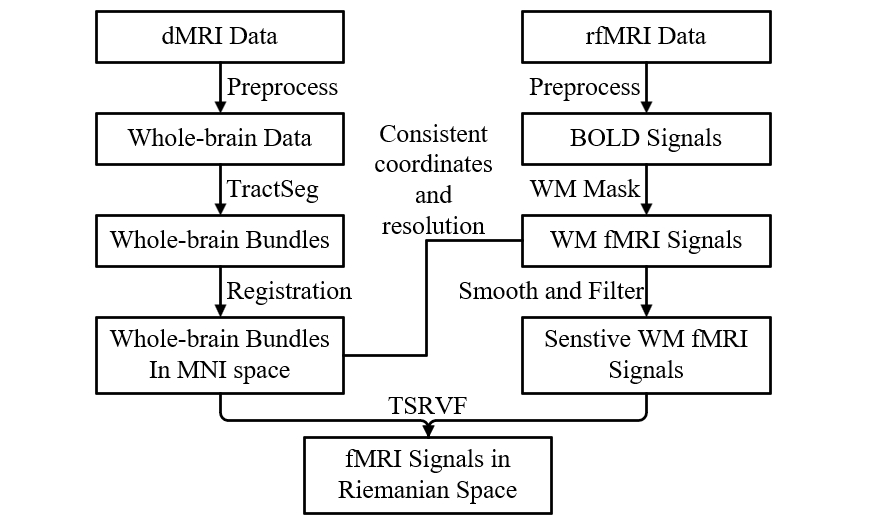}}
%  \vspace{2.0cm}
\end{minipage}
\caption{Flowchart of data preprocessing.}
\label{fig:res1}
\end{figure}

\subsection{Implementation}
\label{ssec:typestyle}
We split the whole dataset into 80 training sets and 20 test sets. In the pretraining stage, our model is trained with an initial learning rate of 3e-3, decaying by a factor of 0.1 every 200 epochs, for a total of 450 epochs. In the fine-tuning stage, the model is trained with a learning rate of 1e-5 for 20 epochs. The batch size is set to 1024, and Adam \cite{Kingma2014-tb} is used for optimization. Deep learning methods were implemented in PyTorch (v2.4.1). To ensure fast and efficient processing of the large number of fiber samples during model training and inference, fibers are downsampled to $n_{p}$ points before being input into the network. In this study, $n_{p}$ is set to 25, as this number provides good performance with relatively low computational costs. 

\subsection{Experimental Results}
\label{ssec:eresults}

To evaluate our clustering results, we compare our method with two state-of-the-art fiber clustering methods: QuickBundles (QB) \cite{Garyfallidis2012-io} and DFC \cite{chen2023deep}. In brief, QB is an efficient fiber clustering technique that groups similar fibers based on extracted features and spatial proximity. DFC is a deep learning-based method, on which our method is built (see Section 2.1). 

We use the following two evaluation metrics. First, as the goal of functional fiber clustering is to ensure that fibers within a cluster exhibit functional homogeneity, we compute the Pearson correlation of fMRI signals at the endpoints of fibers within each cluster \cite{zhao2023riemannian}. A higher correlation indicates a stronger functional correlation across fibers within the cluster. In addition, to assess the geometric similarity of a cluster's fibers, we compute the $\alpha$ measure which is defined as the average pair-wise distances between all fibers within each cluster,  \cite{FFclust}. A lower value of $\alpha$ indicates better clustering performance. 

The average fMRI signal correlation and the $\alpha$ measure for clusters of the 7 bundles of interest is shown in Table 1, where our method in general obtains the best functional performance across all the compared methods and maintains a relatively low $\alpha$ value. These data prove that our clustering method is capable of yielding results with both high functional relevance and geometric consistency.
\begin{table}

\begin{minipage}[b]{1.0\linewidth}
  \centering
  \centerline{\includegraphics[width=9cm]{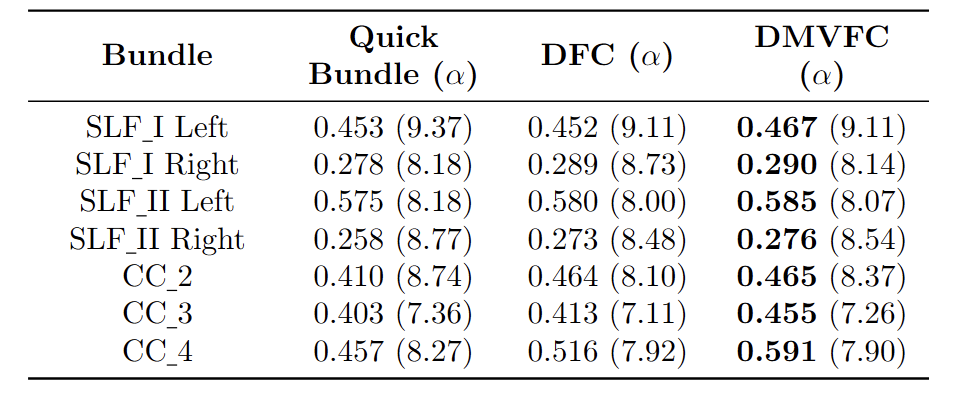}}
%  \vspace{2.0cm}
\end{minipage}
\caption{Comparison of  fMRI signal correlation and $\alpha$ in different bundles.}
\label{fig:res2}
\end{table}

To further demonstrate that our clustering results have functional homogeneity, we visualize the fMRI signals on several example clusters in Figure 3. The colors of the clusters in the figure represent the strength of the fMRI signals. The clusters shown in each column of the figure are taken from the closest spatially located clusters in the same bundle of three different clustering methods, and it can be seen that our method has a more functional coherence compared to the other two methods.

\begin{figure}

\begin{minipage}[b]{1.0\linewidth}
  \centering
  \centerline{\includegraphics[width=8.7cm]{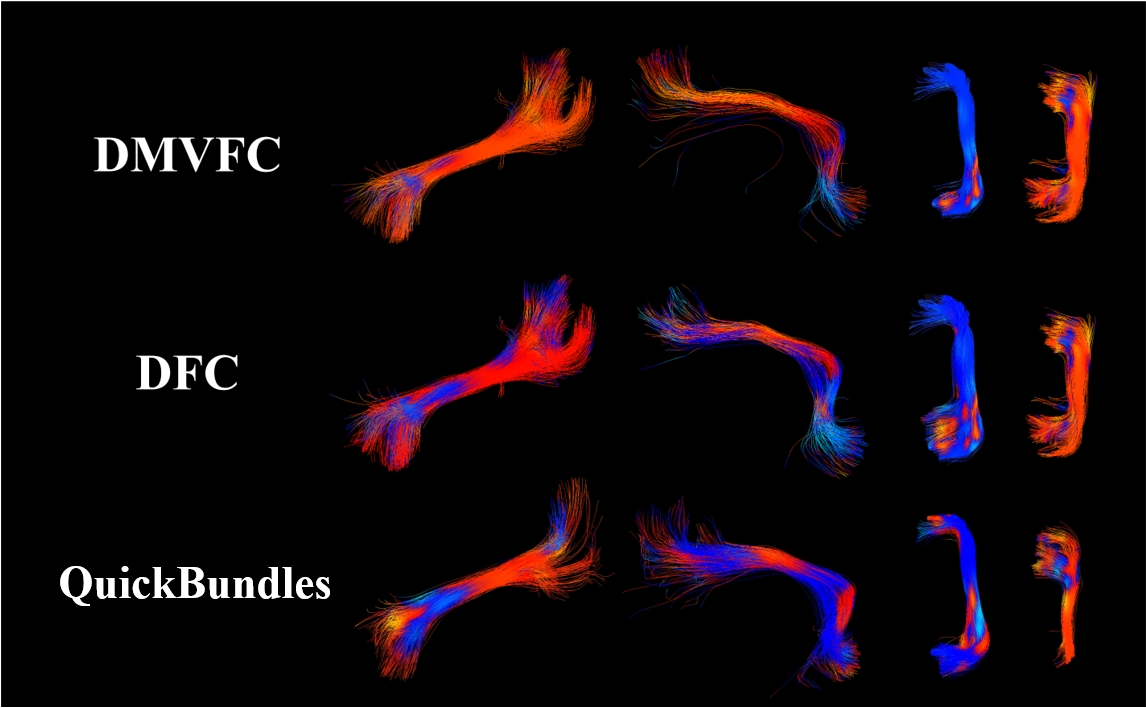}}
%  \vspace{2.0cm}
\end{minipage}
\caption{Example fiber clusters for the compared methods. The regions in red indicate a higher functional coherence, where the signals are similar across the fibers.}
\label{fig:res3}
\end{figure}

\section{Conclusion}
\label{sec:print1}

In this paper, we developed a novel deep-learning framework for tractography fiber clustering that incorporates both geometric and functional information of white matter fiber tracts. The experiments demonstrate the superior performance of our method in achieving functionally meaningful and consistent WM parcellation results. Overall, DMVFC can effectively utilize multimodal dMRI and fMRI information, providing potentially a new framework for fiber clustering techniques.
\newpage
\section{COMPLIANCE WITH ETHICAL STANDARDS}
\label{sec:print}

This study was conducted retrospectively using public HCP imaging data. No ethical approval was required.

\section{Acknowledgments}
\label{sec:acknowledgments}

This work is in part supported by the National Key R\&D Program of China (No. 2023YFE0118600), the National Natural Science Foundation of China (No. 62371107) and the National Institutes of Health (R01MH125860, R01MH119222, R01MH132610, R01NS125781).

%Other types of acknowledgements can also be listed in this section.

%Reporting on real or potential conflicts of interests, or the absence
%thereof, is required in the paper. Authors are responsible for
%correctness of the statements provided in the manuscript. Examples of
%appropriate statements include:
%\begin{itemize}
 % \item ``No funding was received for conducting this study. The
  %  authors have no relevant financial or non-financial interests to
   % disclose.'' 
  %\item ``This work was supported by […] (Grant numbers) and
  %  […]. Author X has served on advisory boards for Company Y.'' 
  %\item ``Author X is partially funded by Y. Author Z is a Founder and
  %  Director for Company C.''
%\end{itemize}


\bibliographystyle{IEEEtran}
\bibliography{strings,refs}

\end{document}